# Conditional Network Analysis Identifies Candidate Regulator Genes in Human B cells

Kai Wang[1,2], Nilanjana Banerjee[2], Adam A. Margolin[1,2], Ilya Nemenman[2], Katia Basso[3], Riccardo Dalla Favera[3], Andrea Califano[1,2,3,*]

[1]Department of Biomedical Informatics, Columbia University, 622 West 168th Street, Vanderbilt Clinic 5th Floor, New York, New York 10032

[2]Joint Centers for Systems Biology, Columbia University, Russ Berrie Pavilion, 1150 St. Nicholas Ave, Rm 121, New York, New York 10032

[3]Institute of Cancer Genetics, Columbia University, Russ Berrie Pavilion, 1150 St. Nicholas Ave., New York, 10032

**ABSTRACT**

Cellular phenotypes are determined by the dynamical activity of networks of co-regulated genes. Elucidating such networks is crucial for the understanding of normal cell physiology as well as for the dissection of complex pathologic phenotypes. Existing methods for such "reverse engineering" of genetic networks from microarray expression data have been successful only in prokaryotes (*E. coli*) and lower eukaryotes (*S. cerevisiae*) with relatively simple genomes. Additionally, they have mostly attempted to reconstruct average properties about the network connectivity without capturing the highly conditional nature of the interactions.

In this paper we extend the *ARACNE* algorithm, which we recently introduced and successfully applied to the reconstruction of whole-genome transcriptional networks from mammalian cells, precisely to link the existence of specific network structures to the expression or lack thereof of specific regulator genes. This is accomplished by analyzing thousands of alternative network topologies generated by constraining the data set on the presence or absence of putative regulator genes. By considering interactions that are consistently supported across several such constraints, we identify many transcriptional interactions that would not have been detectable by the original method. By selecting genes that produce statistically significant changes in network topology, we identify novel candidate regulator genes. Further analysis shows that transcription factors, kinases, phosphatases, and other gene families known to effect biochemical interactions, are significantly overrepresented among the set of candidate regulator genes identified in silico, indirectly supporting the validity of the approach.

# 1 INTRODUCTION

Cellular phenotypes are determined by complex relationships among genes and their products that control the majority of cellular functions. By modeling these complex relationships, the whole genome can be organized into networks of genetic interactions. Understanding this organization is crucial to elucidate normal cell physiology as well as to dissect complex pathologic phenotypes. Over the last few years, a significant effort has been aimed at the systematic reverse-engineering (or deconvolution) of genetic interactions from measurement data, especially microarray expression profiles. Unfortunately, most available methods have been successful only in the study of organisms with relatively simple genomes, such as yeast *Saccharomyces Cerevisiae.*

Recently we introduced a new information-theoretic algorithm, *ARACNE* (Algorithm for the Reconstruction of Accurate Cellular Networks), to reverse-engineer genetic networks from microarray expression profiles (Margolin, Nemenman et al. 2004). *ARACNE* compares favorably with existing reverse-engineering methods, such as Bayesian Networks and Relevance Networks, and scales successfully to large mammalian networks.

*ARACNE* infers interactions based on mutual information between genes, an information-theoretic measure of pairwise correlation. This is not a trivial problem as two genes may be substantially correlated (e.g. because they belong to the same pathway), so that the statistical independence hypothesis $P(g_i, g_j) = P(g_i)P(g_j)$ is rejected with very high probability, without being involved in any direct interaction mechanism (i.e. transcriptional regulation). Given the quadratic explosion of the number of potential pairwise interactions for a fixed number of genes, removing false positive interactions, while still inferring a substantial number of correct ones, is a formidable challenge for all network reconstruction algorithms. We have proven that *ARACNE* can asymptotically reconstruct the exact network structure if the network is a tree. Furthermore, we have shown that it is quite robust to violations of the tree assumption and that it significantly outperforms established network reconstruction methods, such as Bayesian Networks and Relevance Networks, on complex network topologies that contain many tight loops (Margolin, Nemenman et al. 2004).

By applying *ARACNE* to the analysis of expression data from human B cells, we recapitulated a significant number of known targets of the c-MYC proto-oncogene and identified several new ones that were later biochemically validated (Basso, Margolin et al. 2004). In fact, over 90% of the interactions that were tested biochemically (11 of 12) confirmed the in silico inference.

In this paper we continue this line of research and tackle the problem of reconstructing network connectivity not as a static graph, but rather as a large number of graphs conditional on specific molecular constraints. The goal is to identify network configurations that exist in the cell only in the presence or absence of key regulator genes. Such exhaustive analysis is long overdue: even though it is well understood that entire cellular sub-networks are dependent on the expression of key regulator genes, the field has mostly focused on studying static, average information flow among different molecular species in the cell (An important exception is the attempt to couple regulator genes with the modules they control in *S. cerevisiae* (Segal, Shapira et al. 2003)). This makes it hard to recover transient interactions, as well as interactions that cannot be defined in terms of pairwise relationships among genes, such as, for example, a XOR interaction (Margolin, Nemenman et al. 2004; Nemenman 2004).

The genetic network that includes the c-MYC proto-oncogene provides a good example of this type of conditional interactions, as this gene is known to be active as a transcription factor only when it is phosphorylated. BTK and TTK are well-studied kinases that have been shown to mediate cellular response in B cells. Both are essentially statistically independent of c-MYC and are thus not easily identifiable as interacting with the latter. As we will show, however, the analysis of the c-MYC network when these kinases are respectively under- and over-expressed shows a dramatic change in c-MYC related network topology, which is statistically significant and therefore identifies these two kinases, as well as a handful of other genes, as key co-factors or regulators of c-MYC related activities in B cells.

In this paper, we show that by considering each gene as a candidate regulator and by studying the statistical properties of the corresponding conditional networks inferred by *ARACNE*, we can identify a large number of otherwise undetectable pairwise interactions, as well as some three-way interactions that critically determine information flow in cellular networks. *ARACNE* is ideally suited for tasks of this nature since, as shown in Section 1.1 (see also (Margolin, Nemenman et al. 2004)), it already produces very few false positives interactions. Thus studying conserved topological properties across several alternative *ARACNE*-generated networks will further increase the probability of detecting true physical interactions by providing a second, independent statistical filter. Additionally, compared to other algorithms, *ARACNE's* reconstruction is robust even for relatively small sample sizes. Thus, the inference of whole-genome interaction networks based on subsets of the data (constrained by the over or under-expression of a gene) is still feasible.

The proposed method selects all genes with a sufficient dynamic range as potential regulators. For each such gene, it then uses *ARACNE* to reverse-engineer the genetic network using the microarray subsets of in which it is respectively over- and under-expressed. This results in thousands of possible alternative network topologies. We then look for edges that are conserved across many conditions as well as for conditions that produce statistically significant differences in the observed network topology. In Section 3.3 and 3.4, we show that both analyses produce interesting biological results.

**1.1 Background**: *ARACNE* relies on a two-step process. First, candidate interactions are identified by estimating pairwise gene-gene mutual information (MI):

$$I(g_i, g_j) = I[P(g_i, g_j)] = I_{ij} = \left\langle \log \frac{P(g_i, g_j)}{P(g_i)P(g_j)} \right\rangle$$

and by filtering them using an appropriate threshold, $I_0$, computed for a specific p-value, $p_0$, in the null-hypothesis of two independent genes. This step is almost equivalent to the Relevance Networks method (Butte and Kohane 2000), and, correspondingly, suffers from critical limitations. In particular, genes separated by one or more intermediaries may be highly co-regulated without implying a direct physical interaction.

Thus, in its second step, *ARACNE* removes the vast majority of indirect candidate interactions using a well-known property of mutual information – the data processing inequality (DPI) (Cover and Thomas 1991) -- that has not been previously applied to the reverse engineering of networks. The DPI states that if genes $g_1$ and $g_3$ interact only through a third gene, $g_2$, (i.e., if the interaction network is $g_1 \leftrightarrow \ldots \leftrightarrow g_2 \leftrightarrow \ldots \leftrightarrow g_3$ and no alternative path exists between $g_1$ and $g_3$), then the following holds

$$I(g_1, g_3) \leq \min[I(g_1, g_2); I(g_2, g_3)].$$

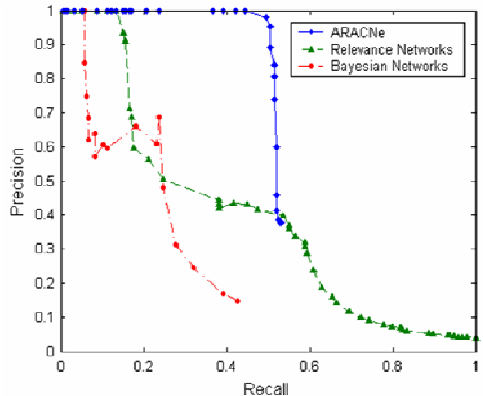

Figure 1: Precision/Recall for 1,000 samples generated from the Mendes scale-free topology. The p-value and the Dirichlet counts were changed respectively in *ARACNE*/RN and BN to produce the Recall Precision Curves.

Correspondingly, *ARACNE* starts with a network graph where each $I_{ij} > I_0$ is represented by an edge $(i, j)$. It then examines each gene triplet, for which all three MIs are greater than $I_0$, and removes the edge with the smallest value. Each triplet is analyzed irrespective of whether one of its edges has been marked for removal by a prior DPI application to a different triplet. Thus the network reconstructed by the algorithm is independent of the order in which the triplets are examined.

As discussed in (Margolin, Nemenman et al. 2004), this allows the exact asymptotic reconstruction of networks with a tree topology, but also performs surprisingly well even if the assumption is significantly violated. For instance, Figure 1 shows analysis of several synthetic networks with 100 genes and 194 interactions using a model proposed by (Mendes, Sha et al. 2003) specifically to benchmark reverse-engineering algorithms. These synthetic networks are modeled using realistic Hill dynamics and have Scale-Free and Erdös-Rényi (random) topologies, thus containing a large number of loops and other complex structures likely to be found in biological networks. Results averaged over three separate network implementations for each topology class are shown in the Figure. We plot *precision* (fraction of true interactions among all predicted by the algorithm) vs. *recall* (fraction of all true interactions uncovered), which are useful performance measures for problems with large potential false positive rate (results are shown only for the scale-free model). As can be seen, *ARACNE* dramatically outperforms both Bayesian Networks (implemented using (Friedman and Elidan 2004)) and Relevance Networks (*ARACNE* without the DPI step) on this data. Note that by multiplying the right-hand side of the DPI by $(1-\tau)$, we can introduce a tolerance which helps reduce the number of false negatives. In (Margolin, Nemenman et al. 2004), we show that values of $\tau < 0.2$ produce an advantageous trade-off between false-negative and false-positives. Choice of $\tau = 1$ corresponds to the Relevance Networks method.

**1.2 Conditional Network Analysis**: As a first attempt to extend the range of interactions reconstructed by *ARACNE* from strictly pairwise to three-way, we propose a systematic approach to study the different network topologies, conditional on a large number of constraints. The latter are based on specific molecular phenotypes. In other words, we study the network $W_k(D | m_k)$ derived from a dataset $D$ conditioned on a molecular phenotype $m_k$. We do this by using the conditional form of the mutual information $I(g_i, g_j | m_k) = I[P(g_i, g_j | m_k)]$ instead of the unconditional one in the *ARACNE* algorithm. Here, by molecular constraint we mean either the up-regulation $g_k^+$ or the down-regulation $g_k^-$ of a specific gene.

Using this approach, we demonstrate the ability of our analysis to discover both a significant number of interactions that would not be detectable in the original version of *ARACNE* as well as key regulator genes that control significant sub-networks, producing a more complete and dynamic picture of the full network topology.

## 2 METHODS

**2.1 Experimental Dataset**: Using the Affymetrix U95A GeneChip System (12,600 probes), we have accumulated an extensive gene expression profile repository for a panel of 336 homogeneous B cell phenotypes derived from normal, tumor-related, and experimentally manipulated populations. This is the same platform that was used for the analysis of the c-MYC centric network in B cells (Basso, Margolin et al. 2004), where *ARACNE* yields a B cell specific regulatory network with approximately 129,000 interactions. Gene-specific subnetworks can be easily extracted by sub-selecting (from the complete network) only those genes that have high mutual information with a particular gene of interest. In particular, we constructed such a sub-network for the proto-oncogene c-MYC, which emerged as one of the 5% largest cellular hubs. For an extensive description of this dataset, please refer to (Klein, Tu et al. 2001).

As proof of concept, we use the same platform to assess the ability of our new analysis to recover new conditional targets of c-MYC, as well as the identity of key regulators or co-factors of this gene. c-MYC constitutes an ideal choice as it is extensively characterized in the literature (Fernandez, Frank et al. 2003; Zeller, Jegga et al. 2003).

**2.2 Candidate Target filtering**: As a first step in our analysis, we pre-select probes that have a mean absolute expression value greater than 50 and standard deviation greater than 30% of the mean as suggested by literature (Golub, Slonim et al. 1999). The reason for this is two-fold: 1) Genes below the threshold do not have enough dynamic range, compared to the error on the expression estimate, to provide useful information. 2) The complexity of *ARACNE* is $O(N^3 + N^2 M^2)$, where *M* is the number of profiles and *N* is the number of probes. Therefore by limiting our data to a more meaningful subset, we speed up the analysis without any significant loss of information.

**2.3 Regulator gene filtering**: In order to identify a reasonable set of potential regulators, we further filter the candidate target set for probes with a distinct expressed vs. unexpressed state. Only probes with a mean expression value greater than 200 and standard deviation greater than 50% of the mean are considered as potential regulators. This eliminates housekeeping genes (expressed at a relatively constant range) and other genes that do not have enough dynamic range to allow us to make a distinction between an over- and an under-expressed state.

We further remove genes with significant mutual information with the hub-gene of interest (e.g. c-MYC). This step is critical. Indeed, imagine selecting a strong repressor of c-MYC as a regulator. This would reduce the dynamic range of c-MYC in the conditional network, not allowing calculating the conditional MI, and the analysis would fail. On the other hand, by definition, conditioning on genes with low mutual information with c-MYC does not change the dynamic range of the latter, allowing the use of *ARACNE*. Note also that elimination of regulators with high mutual information with c-MYC will not lead to extra false negatives, because these dependencies will be identified as direct c-MYC interaction by the regular *ARACNE* method.

Finally, we use hierarchical clustering based on absolute Pearson correlation as the distance measure, so that both correlated and anti-correlated probes are clustered. This produces clusters of probes that are strongly co-regulated across all microarrays and thus correspond to the same molecular constraint. This further reduces the number of potential regulator genes without reducing their ability to constrain the system. These gene clusters will each be investigated as a potential regulator "hyper-gene".

**2.4 The Conditional Adjacency Matrix**: For each candidate regulator genes, $g_k$, we perform two conditional network analyses, by using respectively the set of profiles $G_k^+$ where $g_k$ is expressed (the top 33% of the microarrays rank-ordered by that gene) and the set of profiles $G_k^-$ where it is unexpressed (the bottom 33%). While somewhat arbitrary, this choice usually keeps the two ranges well separated and produces subsets of equal size so that results can be statistically compared. We will consider more sophisticated filtering and range binning methods in the future. For each of these two subsets, *ARACNE* infers an adjacency matrix, where the presence of an interaction between the gene $g_i$ and $g_j$ corresponds to a non-zero entry at the $[i, j]$ position in the matrix.

|  | $G_1^+$ | $G_1^-$ | $G_2^+$ | $G_2^-$ | ... | ... | $G_M^+$ | $G_M^-$ |
|---|---|---|---|---|---|---|---|---|
| Edge 1 | 1 | 0 | 1 | 0 | 0 | 0 | 1 | 0 |
| Edge 2 | 0 | 1 | 0 | 0 | 0 | 1 | 0 | 0 |
| : | 0 | 0 | 1 | 0 | 1 | 0 | 0 | 0 |
| Edge N | 1 | 0 | 0 | 1 | 0 | 0 | 0 | 1 |

Figure 2: The Conditional Adjacency matrix (CA matrix). Columns in the matrix are conditional networks given regulator gene $g_k$ being respectively over-expressed ($G_i^+$) or under-expressed ($G_i^-$). Rows represent interactions with the hub-gene. Ones in the matrix indicate the presence of an interaction.

We collect all the adjacency matrices for all regulators into a single matrix, which we call the Conditional Adjacency Matrix (CA matrix), see Figure 2. The CA matrix is a sparse binary matrix, where columns correspond to candidate regulator genes and their expression states, while rows are the properly ordered network edges, present or not present according to *ARACNE* reconstruction for each condition.

A particularly useful type of conditional adjacency matrix is one where instead of looking at the entire network, we consider only edges incident on a specific gene of interest. Since genes of interest are generally highly connected (genetic hubs), we call this version a hub-specific CA matrix. In the rest of this paper, we consider the hub-specific CA matrix for the c-MYC gene.

**2.5 Network analysis:** By analyzing each row of a hub-specific CA matrix, we identify interactions that are consistently supported (i.e. stable) across multiple regulator constraints. These constitute the "core" of the hub-gene specific subnetwork topology. Similarly, by analyzing columns, we can find conditions that produce changes in network topology that are larger than expected by chance alone.

To produce appropriate null-hypothesis models to assess the statistical significance of our findings, we shuffle the hub-specific CA matrix entries in column-wise fashion for the row-wise analysis and vice versa. Thresholds are then determined using a statistically significant p-value based on such null-hypothesis models.

**2.6 Annotation enrichment analysis**: To analyze the biological relevance of the regulators, we associate each candidate regulator above statistical significance with the Gene Ontology (GO) category (or categories) it belongs to. Categories with fewer than five genes are not considered. For each remaining category, we then calculate the fraction of genes among our candidate regulators and the fraction of genes among all the genes in the GO database. To assign statistical significance to the enrichment of each GO category, we use the hypergeometric distribution. This assesses the probability that the categories are enriched at random. A p-value of $p < 0.01$ is used after correcting for multiple hypotheses using Bonferroni correction.

# 3 RESULTS

Initial filtering of the original gene expression profiles reduces the total number of probes from 12600 to 7484. These probes will be investigated in the conditional network analysis as the pool of candidate c-MYC targets.

**3.1 Candidate regulator gene selection**: Further filtering, see section 2.3, yields 1890 candidate regulator probes with sufficient dynamic range. Of these, only 1109 are statistically independent of c-MYC based on mutual information analysis with a MI threshold corresponding to p-value of $p_0 = 10^{-7}$ or better (yielding less than 5% false positives after correcting for multiple comparisons). Finally, by performing hierarchical clustering of these 1109 candidates with a Pearson correlation threshold of 0.8, we identify 885 unique clusters (including 789 single gene clusters, and up to a single 35-gene cluster). We randomly choose a representative gene from each cluster to generate specific constraints, thus yielding a final list of 885 potential regulators and 1770 specific molecular constraints ($g_k^+, g_k^-$), and microarray subsets ($G_k^+, G_k^-$) associated respectively with the top 33% and bottom 33% of the expression rank of the corresponding potential regulators.

**3.2 Conditional network analysis**: we analyze each microarray subset ($G_k^+, G_k^-$) with *ARACNE* with tolerance $\tau = 0$ to minimize the false positives in each alternative network (see (Margolin, Nemenman et al. 2004) for details). These results are then assembled into a c-MYC specific CA matrix as previously described.

**3.3 Row-wise analysis yields c-MYC first neighbors**: By analyzing the rows in the c-MYC specific CA matrix, we identify edges to c-MYC with a wide spectrum of supports across all conditional networks. The vast majority of the interactions have very low support (i.e. they are identified across fewer than 5 distinct constraints). In order to obtain a null-hypothesis distribution for the edge supports, we independently shuffle each column of the CA matrix.

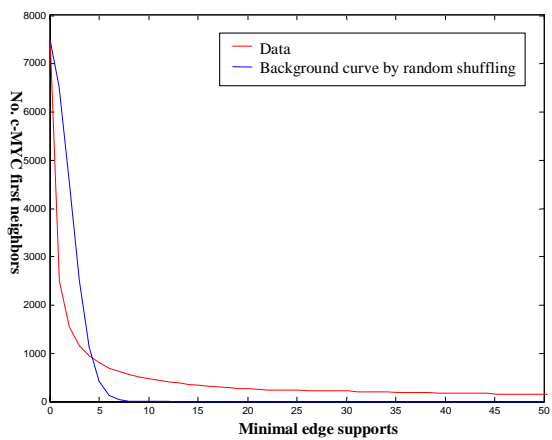

Figure 3: Number of predicted first neighbors of c-MYC as a function of minimal edge supports. The red curve is the number of predicted first neighbors obtained from the dataset by conditional network analysis. By independent shuffling the edges under each conditional network (i.e. the columns of the CA matrix), a null distribution (the blue curve in the figure) is obtained by taking the mean of 100 such trials.

Figure 3 plots the distribution of background edge supports together with the distribution in the analyzed data set.

Table 1 summarizes the number of predicted c-MYC first neighbors for each value of the minimum edge support, as well as the number among these of known c-MYC target genes (Zeller, Jegga et al. 2003), and the expected number of false-positive interactions from the null-hypothesis. We select a minimum support threshold of 9 constraints (i.e. only edges to c-MYC that present in at least 9 conditional networks will be considered statistically significant), yielding fewer than 1% false positives based on the null-hypothesis.

Using this threshold, our conditional network analysis identifies 406 first neighbor of c-MYC, 104 of which have been validated in the c-MYC database (Zeller, Jegga et al. 2003), corresponding to 26% enrichment of known-targets among predicted c-MYC first neighbors (in fact, the enrichment is probably even higher considering that some of the 406 first neighbors of c-MYC may be its transcriptional regulators rather than regulees).

Table 1: c-MYC interaction enrichment ($E$) is calculated as the ratio of validated ($N_V$) and predicted ($N_P$) c-MYC interactions. A p-value ($P$) is associated with each minimal support by computing the ratio of false positive interactions ($N_{FP}$) under the null model and predicted interactions ($N_P$). A minimum support (#) of 9 produces fewer than 1% false positives interactions.

| #  | $N_P$ | $N_V$ | $E$  | $N_{FP}$ | $P$    |
|----|-------|-------|------|----------|--------|
| 1  | 2422  | 437   | 0.18 | 6520.1   | 1      |
| 2  | 1458  | 278   | 0.19 | 4541.5   | 1      |
| 3  | 1066  | 224   | 0.21 | 2514.1   | 1      |
| 4  | 847   | 182   | 0.21 | 1131.6   | 1      |
| 5  | 710   | 157   | 0.22 | 423.13   | 0.60   |
| 6  | 591   | 132   | 0.22 | 136.04   | 0.23   |
| 7  | 511   | 119   | 0.23 | 37.03    | 0.072  |
| 8  | 459   | 110   | 0.24 | 9.18     | 0.02   |
| 9  | 406   | 104   | 0.26 | 1.9      | < 0.01 |
| 10 | 367   | 96    | 0.26 | 0.37     | < 0.01 |
| 11 | 337   | 89    | 0.26 | 0.05     | < 0.01 |
| 12 | 305   | 83    | 0.27 | 0.01     | < 0.01 |
| 13 | 288   | 79    | 0.27 | 0        | < 0.01 |
| 14 | 254   | 72    | 0.28 | 0        | < 0.01 |
| 15 | 232   | 65    | 0.28 | 0        | < 0.01 |
| 16 | 212   | 60    | 0.28 | 0        | < 0.01 |
| 17 | 195   | 58    | 0.30 | 0        | < 0.01 |
| 18 | 179   | 55    | 0.31 | 0        | < 0.01 |
| 19 | 167   | 51    | 0.31 | 0        | < 0.01 |
| 20 | 155   | 48    | 0.31 | 0        | < 0.01 |
| 21 | 143   | 44    | 0.31 | 0        | < 0.01 |
| 22 | 136   | 41    | 0.30 | 0        | < 0.01 |
| 23 | 128   | 40    | 0.31 | 0        | < 0.01 |
| 24 | 125   | 40    | 0.32 | 0        | < 0.01 |
| 25 | 122   | 40    | 0.33 | 0        | < 0.01 |
| 30 | 106   | 35    | 0.33 | 0        | < 0.01 |
| 35 | 83    | 28    | 0.34 | 0        | < 0.01 |
| 40 | 65    | 22    | 0.34 | 0        | < 0.01 |
| 45 | 50    | 18    | 0.36 | 0        | < 0.01 |
| 50 | 44    | 18    | 0.41 | 0        | < 0.01 |

This ratio is statistically significant (p-value $4.86 \times 10^{-21}$) with respect to the expected 11% of background c-MYC targets among randomly selected genes (Fernandez, Frank et al. 2003). Note that, in (Margolin, Nemenman et al. 2004), by applying *ARACNE* without the conditional analysis, we had identified 56 candidate c-MYC interactions of which 22 had been validated in the c-MYC database, corresponding to a 39% enrichment (p-value $4.75 \times 10^{-12}$). Thus, even though the enrichment produced by the conditional analysis method discussed in this paper is lower than the one for the original *ARACNE* method, the p-value is significantly higher because the total number of genes has increased by a factor of seven, reducing the impact of statistical fluctuations. Considering that 90% of previously unreported c-MYC interactions inferred by *ARACNE* were shown to be in fact real targets of c-MYC using Chromatin Immunoprecipitation assays (Basso, Margolin et al. 2004), this indicates that a substantial subset of the 302 previously unreported genes may in fact be bona fide targets.

Additionally, considering that (a) the c-MYC database reports interactions of the gene across a variety of tissues, organisms, and phenotypic conditions, (b) only a fraction of the total genes in the c-MYC database are present in the HU95-A chipset, and only about 60% of the probes have sufficient dynamic range for the analysis, it is likely that the method has recapitulated a substantial subset of all the B cells specific c-MYC interactions that could be effectively measured. As a result, the proposed conditional analysis represents a significant improvement over the original method that inferred only 56 candidate c-MYC target genes.

**3.4 Column-wise analysis yields new regulators of c-MYC behavior**: Each column in the c-MYC specific CA matrix corresponds to a candidate regulator gene either in its over or under-expressed state. We hypothesize that candidate regulator genes that produce the largest change in c-MYC connectivity (either for their over- or under-expressed range) will be important biological co-regulators of c-MYC specific functions.

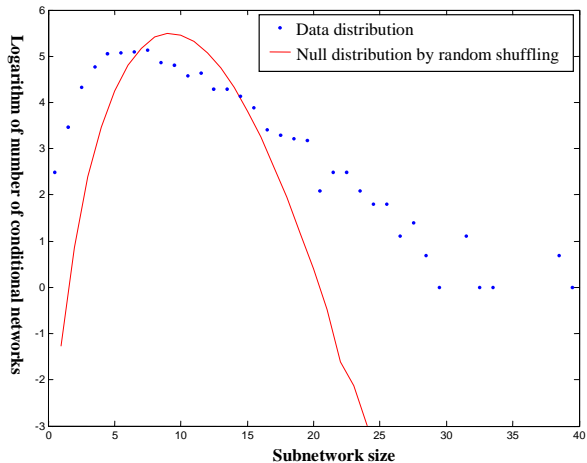

Figure 4: Distribution of the size of conditional networks. The size of a c-MYC specific subnetwork is defined as the number of predicted first neighbors of c-MYC. Plotted on the Y-axis is number of conditional networks in logarithm scale. The blue dots are the subnetwork size distribution measured from data. A background size distribution is obtained by independently shuffling the edges across all conditional networks (i.e. the rows of the CA matrix). The mean of 100 such trials is plotted in the figure as the red curve.

To validate this hypothesis, we first produce null-hypothesis statistics by independently shuffling the entries in the CA matrix across each row. The background distribution and the distribution obtained from our analysis are plotted in Figure 4. A threshold of 18 edges yields fewer than 5% expected false prediction of key regulator genes (i.e., only constraints that produce at least 18 c-MYC interactions will be considered statistically significant). This produces a list of 118 unique candidate regulator genes whose presence/absence results in much larger variations of c-MYC connectivity than could be expected by chance alone. After including all the genes in the cluster of each regulator gene (105 of them are single gene clusters from the hierarchical clustering step), we produce a final list of 168 candidate regulators.

**3.5 Annotation enrichment analysis of modulator genes**: We evaluate the GO category enrichment for the set of statistically significant candidate regulators for c-MYC. Table 2 summarizes the enriched categories, including the p-value corrected for multiple testing across multiple GO categories. The set of 168 candidate regulators contains 116 GO annotated genes that are significantly enriched for molecular functions and biological process categories that are highly relevant to cellular control, particularly in B cells. 21 out of 116 regulator genes are involved in transcription regulation activity and 13 are protein kinases, including many of the previously-characterized kinases important for B cell e.g. BTK, TTK, IKBKE. In addition, the regulator genes are also enriched in immune response and humoral response processes. The latter is not surprising given the nature and activity of B cells. In general, the uncharacterized regulators now provide focused hypotheses about their role in the c-MYC-centered network.

## 4 CONCLUSION AND DISCUSSION

Using the *ARACNE* platform, we have performed conditional network analysis on a panel of 336 gene expression profiles from human B-cell populations. This approach is novel in that it combines conditional analysis of molecular phenotypes with the reconstruction of gene regulatory networks. The analysis focuses on the reconstruction of a hub-gene specific (e.g. c-MYC) subnetwork under the constraint of genes (potential regulators) that do not appear to directly interact with the hub-gene. This allows us to identify interactions with the hub-gene that are not supported across all samples, but only detectable conditional on the over- or under-expression of another gene. Also based on this approach, we can identify potential regulators of the hub-gene specific subnetwork that significantly change the topology of the network around the hub.

We have reversed-engineer the subnetwork that includes c-MYC, an important proto-oncogene implicated in lymphomagenesis, from a large set of B cell specific microarray expression profiles.

Compared with the global network analysis, which predicts 56 interactions, 22 of which are validated in the literature, the conditional network analysis predicts 406 interactions, 104 of which are validated in the literature. While the enrichment of validated c-MYC interactions among the predicted ones drops about 10% (from 39% to 26%), the new results has a much higher p-value, based on the expected background of c-MYC interactions, because of the much larger number of candidate interactions (a seven-fold increase with respect to the original *ARACNE* work). We believe that this constitutes a significant improvement in the methodology by allowing the identification of a significantly larger number of candidate interactions without increasing significantly the number of false positives. Additionally, given the results of previous biochemical validation results on *ARACNE*'s prediction, it is reasonable to assume that the actual number of c-MYC target genes, among the predicted ones, is much larger than the 106 validated ones from the literature and may be as much as 90%.

Finally we show that the candidate regulators are highly enriched for molecular functions and biological processes that are consistent with the modulation of the behavior of other genes, either transcriptionally or post-translationally. A much larger number than expected by chance of kinases (which may be involved in the phosphorylation of c-MYC) and transcription factors (which may be c-MYC co-factors) were identified among the candidate regulator. These include some well-known genes, such as the kinases BTK, TTK and IKBKE, which are known to play an important role in B cell physiology.

The improvements resulting from this research may come from two sources: 1) we believe the conditional analysis is able to identify interactions that are only present in a subset of the entire samples, hence are diluted if the analysis is done globally. These interactions with c-MYC can be observed only when conditional on a third gene that may be co-factors of c-MYC, or post-translationally modifies c-MYC. 2) In the global analysis the DPI may remove some of the interactions that are indeed presented in the network. With the conditional analysis, we can recover some of these true interactions in a constrained subset of samples where these interactions become manifested.

Some key differences of the Bayesian Networks (BN) approach (Segal, Shapira et al. 2003) with this method are: (a) BN explicitly select a list of candidate regulator genes, which removes the exponentially long search though the space of network topologies. Our method considers every reasonably expressed gene as a potential regulator (b) BN bin both regulator and target gene dynamics while our method only bins the expression of the potential regulators, and (c) *ARACNE* produces consistently better results than BN from comparative benchmarks, see Section 1.1.

Table 2: GO category enrichment for candidate regulator genes. MF corresponds to Molecular Function and BP corresponds to Biological Process. Columns are respectively the number of candidate regulators in a specific category ($N_C$), the total number of GO annotated candidate regulators from our analysis ($N_{Tot}$), the number of genes in the GO category (GO $N_C$), the number of total genes across all categories (different for MF and BP) (GO $N_{Tot}$), and the p-value (*P*)

| GO Category | $N_C$ | $N_{Tot}$ | GO $N_C$ | GO $N_{Tot}$ | *P* |
|---|---|---|---|---|---|
| Transcription Regulator Activity (MF) | 21 | 116 | 2089 | 21014 | 0.0049 |
| Protein Kinase (MF) | 13 | 116 | 1004 | 21014 | 0.003 |
| IKBK/NFKB cascade (BP) | 5 | 117 | 222 | 24373 | 0.004 |
| Immune Response (BP) | 19 | 117 | 1664 | 24373 | 0.0003 |
| Humoral IR (BP) | 9 | 117 | 378 | 24373 | 0.0001 |
| Reg. of Transcription, DNA-dep. (BP) | 26 | 117 | 1697 | 24373 | $1.2 \times 10^{-7}$ |